\documentclass[summary]{URSIGASS2020}

\usepackage[per-mode=symbol]{siunitx}
\usepackage{cite}
\usepackage{tikz}
\usepackage{xcolor}
\usepackage{pgfplots}
\usepackage{pgfplotstable}
\usetikzlibrary{arrows}
\pgfplotsset{compat=newest}

\definecolor{TUMBlue}{RGB/cmyk}{0, 101, 189/1.00, 0.43, 0.00, 0.00}
\definecolor{TUMWhite}{RGB/cmyk}{255, 255, 255/0.00, 0.00, 0.00, 0.00}
\definecolor{TUMBlack}{RGB/cmyk}{0, 0, 0/0.00, 0.00, 0.00, 1.00}
\definecolor{TUMBlueDark}{RGB/cmyk}{0, 82, 147/1.00, 0.54, 0.04, 0.19}
\definecolor{TUMBlueDarker}{RGB/cmyk}{0, 51, 89/1.00, 0.57, 0.12, 0.70}
\definecolor{TUMGrayDark}{RGB/cmyk}{88, 88, 90/0.00, 0.00, 0.00, 0.80}
\definecolor{TUMGray}{RGB/cmyk}{156, 157, 159/0.00, 0.00, 0.00, 0.50}
\definecolor{TUMGrayLight}{RGB/cmyk}{217, 218, 219/0.00, 0.00, 0.00, 0.20}
\definecolor{TUMGreen}{RGB/cmyk}{162, 173, 0/0.35, 0.00, 1.00, 0.20}
\definecolor{TUMOrange}{RGB/cmyk}{227, 114, 34/0.00, 0.65, 0.95, 0.00}
\definecolor{TUMIvory}{RGB/cmyk}{218, 215, 203/0.03, 0.04, 0.14, 0.08}
\definecolor{TUMBlueLight}{RGB/cmyk}{100, 160, 200/0.65, 0.19, 0.01, 0.04}
\definecolor{TUMBlueLighter}{RGB/cmyk}{152, 198, 234/0.42, 0.09, 0.00, 0.00}
\definecolor{TUMExtViolet}{RGB/cmyk}{105, 8, 90/0.50, 1.00, 0.00, 0.40}
\definecolor{TUMExtNavy}{RGB/cmyk}{15, 27, 95/1.00, 1.00, 0.00, 0.40}
\definecolor{TUMExtTeal}{RGB/cmyk}{0, 119, 138/1.00, 0.03, 0.30, 0.30}
\definecolor{TUMExtForest}{RGB/cmyk}{0, 124, 48/1.00, 0.00, 1.00, 0.20}
\definecolor{TUMExtLime}{RGB/cmyk}{103, 154, 29/0.60, 0.00, 1.00, 0.20}
\definecolor{TUMExtYellow}{RGB/cmyk}{255, 220, 0/0.00, 0.10, 1.00, 0.00}
\definecolor{TUMExtGoldenrod}{RGB/cmyk}{249, 186, 0/0.00, 0.30, 1.00, 0.00}
\definecolor{TUMExtPumpkin}{RGB/cmyk}{214, 76, 19/0.00, 0.80, 1.00, 0.10}
\definecolor{TUMExtRed}{RGB/cmyk}{196, 7, 27/0.10, 1.00, 1.00, 0.10}
\definecolor{TUMExtMaroon}{RGB/cmyk}{156, 13, 22/0.00, 1.00, 1.00, 0.40}

\title{Monte Carlo Modeling of Terahertz Quantum Cascade Detectors}

\author{Johannes Popp*\affref{ref1}, Michael Haider\affref{ref1},
  Martin Francki\'{e}\affref{ref2},  J\'{e}r\^{o}me Faist\affref{ref2}
  and Christian Jirauschek\affref{ref1}}

\affiliation{%
  \aff{ref1}{Department of Electrical and Computer Engineering, Technical
    University of Munich,\\
    Arcisstr. 21, 80333 Munich, Germany}
  \aff{ref2}{Department of Physics, ETH Zurich, Auguste-Piccard-Hof 1, 8093 Zurich, Switzerland}
}

\begin{document}
  
\maketitle

\begin{abstract}
We demonstrate an Ensemble Monte Carlo (EMC) modeling approach for robust and rigorous simulations of  photovoltaic quantum cascade detectors (QCDs) in the mid-infrared (mid-IR) and terahertz (THz) range. The existing EMC simulation tool for quantum cascade lasers (QCLs) was extended to simulate the photovoltaic transport effects in QCDs at thermal equilibrium under zero bias. Here, we present the results of the EMC study of a THz detector design with a detection wavelength of \SI{84}{\micro\meter}. The simulation results show good agreement with experimental data. For a temperature of \SI{10}{\kelvin} we obtain a peak responsivity of \SI{9.4}{\milli\ampere\per\watt}.
\end{abstract}

\section{Introduction}
The detection of light in the mid-IR and THz regime can be accomplished based on intersubband transitions (ISB) in quantum well (QW) structures, where a suitable optical transition between quantized levels in the conduction band of a semiconductor heterostructure is utilized. ISB photodetectors are divided into two groups depending on their operating regime, quantum well infrared photodetectors (QWIPs) and quantum cascade detectors (QCDs).
The first realized and most common ISB detector design is the QWIP\cite{levine1987}. Electrons are excited by photon absorption from the ground level to the upper level in the active well and contribute to the photocurrent under an external bias. Common QWIP designs are based on bound-to-miniband \cite{bandara1992gaas}, bound-to-bound \cite{levine1987} and bound-to-quasi-bound transitions \cite{liu1993dependence}. Alternatively, photovoltaic ISB photodetectors can be utilized, which are commonly based on the design of quantum cascade lasers (QCLs) and are called quantum cascade detectors (QCDs) \cite{hofstetter2002, gendron2004}. The asymmetric conduction band potential makes quantum cascade structures feasible for the usage as a photodetector without applying an external electric field. The major advantage of QCDs in comparison to QWIPs is the absence of dark current noise. For QCDs, the dominating noise contribution is thus given by Johnson noise. Furthermore QCDs offer increased design freedom in choice of material system and compatibility to QCL fabrication technology. Different designs with vertical \cite{giorgetta2009} and diagonal \cite{reininger2014} absorbing transitions have been realized. Recently published detectors are based on a coupled quantum well design\cite{dougakiuchi2016} and can be used as  bi-functional quantum cascade laser and detector (QCLD) devices enabling the on-chip integration for gas sensing\cite{schwarz2012}. 

\begin{figure}
  \centering
  \input{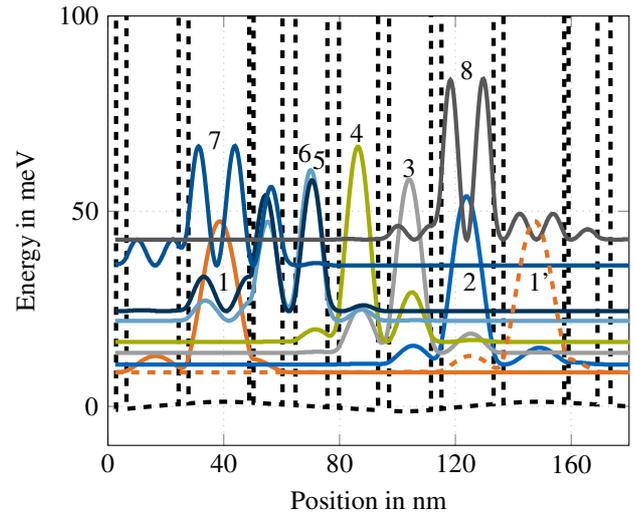}
  \caption{Calculated conduction band profile and probability densities of the investigated THz QCD structure \cite{graf2004} at 10 K. Space charge effects are included by solving self-consistently the Schr\"odinger and Poisson equation. The electron densities in each state are modeled accounting for thermal equilibrium under zero external bias and no incident light.}
  \label{fig:wavefunctions}
\end{figure}

The first THz QCD design was published by Graf \textit{et al.} for a detection wavelength of 
$\SI{84}{\micro\meter}$\cite{graf2004}. The bandstructure and calculated wavefunctions are presented in Fig.~\ref{fig:wavefunctions}.
The QCL structure consists of multiple periods of a doped active QW and an adjacent extraction cascade of QWs with varying thicknesses. The QWs are based on an AlGaAs/GaAs  material system. Photo-exitation occurs between the ground level $1$ and the upper levels $5,6$ followed by the extraction through the staircase of subbands (levels 4, 3 and 2) to the ground state 1' of the adjacent period. A second possible path of relaxation is the back-hopping to ground level 1 which results in zero net contribution to the photocurrent. 

In QWIPs the electron transport between quantized electron states in confined wells and 3D electronic states in the continuum is difficult to model.  In contrast, the well-established simulation models for QCLs (e.g., EMC \cite{jirauschek2014}) can be adapted to QCDs by some minor extensions \cite{baumgartner2013, harrer2016}. Detectors in photovoltaic mode operate close to thermal equilibrium where the detailed balance principle holds, which is usually not strictly fulfilled in na\"ive numerical implementations \cite{frensley1990boundary}. We find that this issue necessitates an adaptation of the EMC simulation, as described in the next section. We present the implementation of necessary extensions of our EMC modeling tool for the carrier transport simulations of QCDs and discuss the obtained results for the given THz QCL structure. 
\section{QCD modeling}
For the characterization of photodetector performance a key figure of merit is the responsivity $R$, which is defined by the generated detector photocurrent $I_\mathrm{out}$ per incident optical power $P_\mathrm{in}$.  
The frequency dependent responsivity can be expressed as \cite{harrer2016}
\begin{equation}
R(\nu) =  \frac{I_\mathrm{out}}{P_\mathrm{in}} =\frac{e}{\hbar \omega}\frac{p_\mathrm{e}}{N_\mathrm{p}} T_\mathrm{f}\lbrack 1 - \exp( -\alpha N_\mathrm{p} L_\mathrm{p} \sin \theta) \rbrack,
\end{equation}
 where $T_\mathrm{f}$ is the facet reflectivity, $N_\mathrm{p}$ the number of periods in the active region, $L_\mathrm{p}$ the length of one period, $e$ the elementary charge, $\omega = 2 \pi c/\lambda$ the angular frequency and $\hbar$ the reduced Planck constant. The propagation angle $\theta$ of the light relative to the growth direction is \SI{45}{\degree} for  the modeled mesa device. The extraction efficiency $p_\mathrm{e}$ and the absorption coefficient $\alpha$ can be calculated using the EMC simulation results\cite{jirauschek2014, jirauschek2017density}.
 
The extraction efficiency $p_\mathrm{e}$ represents the ratio of photoexited electrons scattering to the ground level of the next period to absolute number of photoexcited electrons. In our approach, $p_\mathrm{e}$ is calculated based on the scattering rates extracted from the EMC simulations. Here we assume periodic boundaries and use the scattering rates between states in two adjacent periods to calculate the extraction efficiency. We have developed a robust method where $p_\mathrm{e}$ is directly extracted from the absorption-induced net current, obtained by solving perturbed rate equations in analogy to \cite{jirauschek2018universal}.
%The stationary solution of the rate equation for a QCD is than given by 
% \begin{equation}
% Q \mathbf{p} = \mathbf{b},
% \end{equation}
% with the transition rate matrix $Q$ and the subband occupation vector $\mathbf{p}$. The  diagonal matrix elements are given by $q_{i,i} = -r_i$ and the offdiagonal elements by $q_{i,j} = r_{j+N,i} + r_{j,i} + r_{j,i+N}$ for $i = 1 .. (N-1)$, and $q_{N,j}= 1$  for $j = 1 .. N$. The vector $\mathbf{b}$ has one non-zero element $b_N= 1$. By adding two additional terms $\delta q_{g,u} = r_p$ and $\delta q_{g,g} = r_p$ accounting for the absorbing transition from ground level $g$ to upper level $u$ in case of illumination a change in subband occupation  $\delta \mathbf{p}$ results.
%The extraction efficiency can be determined with this artifical pertubation by calculating the change in net current leaving one period to change in current from ground to upper level:
%\begin{equation}
%p_\mathrm{e} = \frac{\sum_{i = 1}^{N}\sum_{j = 1}^{N} (r_{i,j+N}-r_{i,j-N})\delta p_i}{p_g \delta q_{g,u}}.
%\end{equation}
 
The simulation of photovoltaic QCDs differ from that of QCLs. Here we assume a thermal distribution of charges under zero bias and without illumination. The electron sheet density in a subband $i$ at temperature $T$ is then calculated using the 2D density of states $n^\mathrm{2D}_i= m_i/(\pi\hbar^2)$:
\begin{equation}
n_{\mathrm{s},i}= \frac{m_i}{\pi\hbar^2}k_\mathrm{B}T \ln \lbrace 1 + \exp\lbrack(\mu -E_i)/(k_\mathrm{B} T)\rbrack \rbrace.
\label{eq:occ}
\end{equation}
In Eq.~\ref{eq:occ}, $E_i$ is the quantized subband energy, $m_i$ is the effective mass containing non-parabolicity effects and $\mu$ the chemical potential. Furthermore, the principle of detailed balance holds for the scattering rates between each pair of subbands $i$ and $j$ \cite{koeniguer2006,iotti2005}, 
\begin{equation}
p_i r_{i,j} = p_j r_{j,i},
\end{equation}
with the subband occupation probabilities $p_i$ and $p_j$.
For the determination of scattering rates we use the EMC simulation model\cite{jirauschek2014}, where the scattering is self-consistently evaluated based on Fermi's golden rule. We account for optical and acoustic phonons, interface roughness, impurity and alloy disorder scattering mechanisms. The wavefunctions and eigenenergies necessary for calculating the scattering form factors are identified by a Schr\"odinger-Poisson solver \cite{jirauschek2009accuracy}. 

\section{Results}
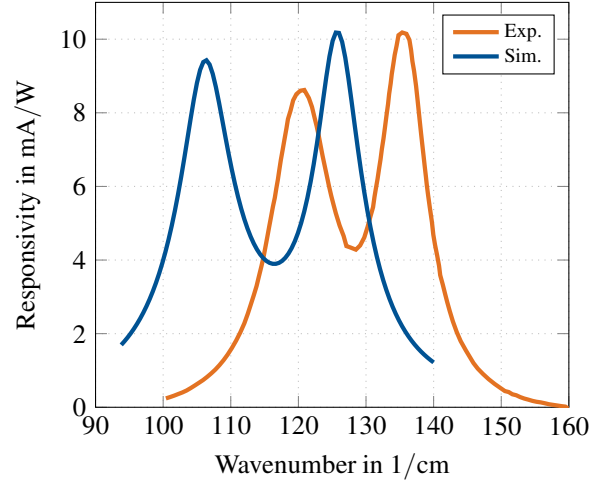
\begin{figure}
  \centering
  \begin{tikzpicture}

    \begin{axis}[%
      width=0.75\columnwidth,
      scale only axis,
      %area style,
      xmin=90,
      xmax=160,
      xlabel={Wavenumber in $\si{\per\centi\meter}$},
      %separate axis lines,
      %every outer y axis line/.append style={black},
      %every y tick label/.append style={font=\color{black}},
      ymin=0,
      ymax=11,
      scaled ticks=false,
      ylabel={Responsivity in $\si{\milli\ampere\per\watt}$},
      %y tick label style={
      %    /pgf/number format/.cd,
      %    fixed,
      %    fixed zerofill,
      %    precision=2,
      %    /tikz/.cd
      %  },
      grid=both,
      grid style={draw=gray!50, dotted},
      legend style={legend cell align=left,align=left,nodes={scale=0.75, transform shape}},
      legend pos=north east,
    ]
    
    %\addplot [TUMBlack,thick,solid] table [y=E3ideal, x=time] {\datatable};
    \addlegendentry{Exp.};
    \addlegendentry{Sim.};
    
    \addplot [TUMOrange,ultra thick]
    table[x=k,  y=R]
    {%
         k	R
         1.004060e+02	2.408994e-01
         1.010157e+02	2.768309e-01
         1.016253e+02	3.158181e-01
         1.022349e+02	3.598982e-01
         1.028446e+02	4.029597e-01
         1.034542e+02	4.546790e-01
         1.040638e+02	5.201491e-01
         1.046734e+02	5.907120e-01
         1.052830e+02	6.638213e-01
         1.058925e+02	7.379492e-01
         1.065021e+02	8.253185e-01
         1.071116e+02	9.203270e-01
         1.077212e+02	1.024503e+00
         1.083307e+02	1.145485e+00
         1.089402e+02	1.280727e+00
         1.095496e+02	1.434303e+00
         1.101591e+02	1.608760e+00
         1.107936e+02	1.811728e+00
         1.114115e+02	2.034741e+00
         1.120112e+02	2.282472e+00
         1.125146e+02	2.539217e+00
         1.129673e+02	2.795465e+00
         1.134066e+02	3.046276e+00
         1.138146e+02	3.292618e+00
         1.141529e+02	3.548944e+00
         1.145343e+02	3.831923e+00
         1.148599e+02	4.094321e+00
         1.151397e+02	4.369132e+00
         1.154051e+02	4.629066e+00
         1.156758e+02	4.901135e+00
         1.159525e+02	5.171334e+00
         1.162292e+02	5.447136e+00
         1.165156e+02	5.693802e+00
         1.167362e+02	5.946936e+00
         1.169657e+02	6.185128e+00
         1.171788e+02	6.449362e+00
         1.174015e+02	6.716981e+00
         1.176788e+02	6.999126e+00
         1.179555e+02	7.291651e+00
         1.182385e+02	7.582197e+00
         1.184554e+02	7.828610e+00
         1.188386e+02	8.058777e+00
         1.192262e+02	8.339361e+00
         1.197066e+02	8.506561e+00
         1.202912e+02	8.600000e+00
         1.208778e+02	8.619473e+00
         1.215455e+02	8.401674e+00
         1.220101e+02	8.159676e+00
         1.223764e+02	7.878685e+00
         1.227093e+02	7.632067e+00
         1.230090e+02	7.376498e+00
         1.233005e+02	7.092922e+00
         1.235780e+02	6.829518e+00
         1.238556e+02	6.588523e+00
         1.241636e+02	6.322866e+00
         1.245188e+02	6.013496e+00
         1.248518e+02	5.748951e+00
         1.252291e+02	5.481189e+00
         1.255343e+02	5.271395e+00
         1.260947e+02	4.937299e+00
         1.266874e+02	4.654730e+00
         1.271031e+02	4.378825e+00
         1.276180e+02	4.346639e+00
         1.285012e+02	4.282275e+00
         1.291315e+02	4.396119e+00
         1.297442e+02	4.612181e+00
         1.302434e+02	4.904937e+00
         1.306615e+02	5.188102e+00
         1.310426e+02	5.436083e+00
         1.312585e+02	5.720629e+00
         1.314798e+02	5.963839e+00
         1.317093e+02	6.205018e+00
         1.319306e+02	6.481307e+00
         1.321519e+02	6.741163e+00
         1.323473e+02	7.018956e+00
         1.325850e+02	7.298051e+00
         1.328063e+02	7.606415e+00
         1.330275e+02	7.925324e+00
         1.332583e+02	8.233438e+00
         1.334197e+02	8.511283e+00
         1.336454e+02	8.793506e+00
         1.338068e+02	9.048943e+00
         1.340323e+02	9.312959e+00
         1.342415e+02	9.570579e+00
         1.345243e+02	9.821581e+00
         1.348073e+02	1.010817e+01
         1.353530e+02	1.018160e+01
         1.359746e+02	1.014542e+01
         1.364183e+02	9.979842e+00
         1.366682e+02	9.698522e+00
         1.368627e+02	9.398335e+00
         1.371070e+02	9.137025e+00
         1.374014e+02	8.731320e+00
         1.376126e+02	8.370464e+00
         1.378071e+02	8.026794e+00
         1.379738e+02	7.731686e+00
         1.381406e+02	7.440313e+00
         1.383184e+02	7.146695e+00
         1.385111e+02	6.831410e+00
         1.386852e+02	6.541528e+00
         1.388519e+02	6.277045e+00
         1.390020e+02	6.020785e+00
         1.391742e+02	5.774974e+00
         1.394344e+02	5.427865e+00
         1.397518e+02	5.027594e+00
         1.399519e+02	4.690111e+00
         1.402073e+02	4.412150e+00
         1.405292e+02	4.116237e+00
         1.408068e+02	3.838268e+00
         1.409735e+02	3.588724e+00
         1.413619e+02	3.327147e+00
         1.416837e+02	3.089496e+00
         1.420000e+02	2.863799e+00
         1.423884e+02	2.609690e+00
         1.428489e+02	2.331653e+00
         1.434147e+02	2.044612e+00
         1.439931e+02	1.799183e+00
         1.447596e+02	1.547201e+00
         1.458872e+02	1.202249e+00
         1.465281e+02	1.049952e+00
         1.473001e+02	9.048084e-01
         1.481315e+02	7.644446e-01
         1.490216e+02	6.346041e-01
         1.499880e+02	5.099747e-01
         1.505369e+02	4.403033e-01
         1.510967e+02	4.143244e-01
         1.516456e+02	3.530562e-01
         1.522172e+02	3.267528e-01
         1.527979e+02	2.769171e-01
         1.533945e+02	2.374066e-01
         1.540042e+02	2.071311e-01
         1.546140e+02	1.737998e-01
         1.552237e+02	1.491265e-01
         1.558334e+02	1.331109e-01
         1.564432e+02	1.094561e-01
         1.570529e+02	8.223630e-02
         1.576627e+02	6.520219e-02
         1.582724e+02	4.867736e-02
         1.588821e+02	3.113397e-02
         1.594918e+02	1.766485e-02
         1.598891e+02	-3.358175e-02
    };
    
    \addplot [TUMBlueDark, ultra thick]
    table[x=k,  y=R]
    {%
         k	R
         9.380000e+01	1.694043e+00
         9.426667e+01	1.790204e+00
         9.473333e+01	1.894411e+00
         9.520000e+01	2.007520e+00
         9.566667e+01	2.130493e+00
         9.613333e+01	2.264412e+00
         9.660000e+01	2.410491e+00
         9.706667e+01	2.570097e+00
         9.753333e+01	2.744760e+00
         9.800000e+01	2.936195e+00
         9.846667e+01	3.146314e+00
         9.893333e+01	3.377233e+00
         9.940000e+01	3.631281e+00
         9.986667e+01	3.910972e+00
         1.003333e+02	4.218960e+00
         1.008000e+02	4.557937e+00
         1.012667e+02	4.930437e+00
         1.017333e+02	5.338496e+00
         1.022000e+02	5.783085e+00
         1.026667e+02	6.263197e+00
         1.031333e+02	6.774467e+00
         1.036000e+02	7.307233e+00
         1.040667e+02	7.844197e+00
         1.045333e+02	8.358388e+00
         1.050000e+02	8.813065e+00
         1.054667e+02	9.165849e+00
         1.059333e+02	9.378154e+00
         1.064000e+02	9.426994e+00
         1.068667e+02	9.312659e+00
         1.073333e+02	9.057501e+00
         1.078000e+02	8.697320e+00
         1.082667e+02	8.270980e+00
         1.087333e+02	7.812890e+00
         1.092000e+02	7.349625e+00
         1.096667e+02	6.899672e+00
         1.101333e+02	6.474730e+00
         1.106000e+02	6.081436e+00
         1.110667e+02	5.722954e+00
         1.115333e+02	5.400209e+00
         1.120000e+02	5.112780e+00
         1.124667e+02	4.859511e+00
         1.129333e+02	4.638906e+00
         1.134000e+02	4.449389e+00
         1.138667e+02	4.289466e+00
         1.143333e+02	4.157822e+00
         1.148000e+02	4.053389e+00
         1.152667e+02	3.975386e+00
         1.157333e+02	3.923342e+00
         1.162000e+02	3.897128e+00
         1.166667e+02	3.896972e+00
         1.171333e+02	3.923489e+00
         1.176000e+02	3.977706e+00
         1.180667e+02	4.061090e+00
         1.185333e+02	4.175590e+00
         1.190000e+02	4.323663e+00
         1.194667e+02	4.508304e+00
         1.199333e+02	4.733052e+00
         1.204000e+02	5.001945e+00
         1.208667e+02	5.319393e+00
         1.213333e+02	5.689868e+00
         1.218000e+02	6.117297e+00
         1.222667e+02	6.603913e+00
         1.227333e+02	7.148252e+00
         1.232000e+02	7.741917e+00
         1.236667e+02	8.364972e+00
         1.241333e+02	8.980858e+00
         1.246000e+02	9.533940e+00
         1.250667e+02	9.955175e+00
         1.255333e+02	1.017953e+01
         1.260000e+02	1.016926e+01
         1.264667e+02	9.928206e+00
         1.269333e+02	9.497347e+00
         1.274000e+02	8.936783e+00
         1.278667e+02	8.306734e+00
         1.283333e+02	7.655964e+00
         1.288000e+02	7.018371e+00
         1.292667e+02	6.414565e+00
         1.297333e+02	5.855230e+00
         1.302000e+02	5.344416e+00
         1.306667e+02	4.882106e+00
         1.311333e+02	4.465998e+00
         1.316000e+02	4.092641e+00
         1.320667e+02	3.758131e+00
         1.325333e+02	3.458519e+00
         1.330000e+02	3.190031e+00
         1.334667e+02	2.949175e+00
         1.339333e+02	2.732787e+00
         1.344000e+02	2.538039e+00
         1.348667e+02	2.362423e+00
         1.353333e+02	2.203729e+00
         1.358000e+02	2.060016e+00
         1.362667e+02	1.929580e+00
         1.367333e+02	1.810931e+00
         1.372000e+02	1.702765e+00
         1.376667e+02	1.603937e+00
         1.381333e+02	1.513445e+00
         1.386000e+02	1.430411e+00
         1.390667e+02	1.354061e+00
         1.395333e+02	1.283716e+00
         1.400000e+02	1.218776e+00
         };
    
    \end{axis}
    
    \end{tikzpicture}%
  \caption{Simulated THz QCD reponsivity (blue line) and measured detector response (orange line) given by Graf \textit{et al.}\cite{graf2004} at the temperature of \SI{10}{\kelvin}. The responsibity simulations are carried out in the linear regime. Results are based on the simulated absorption efficiency of a \SI{45}{\degree} mesa device used for the experimental measurement.}
  \label{fig:responsivity}
\end{figure}
The simulated results for the responsivity together with the experimental results from \cite{graf2004} are shown in Fig.~\ref{fig:responsivity}. The two peaks correspond to the transitions from the ground level $1$ to the upper levels $5$ and $6$, respectively. Here, the calculated energy values for the diagonal transitions from the ground level 1 up to the levels 5 and 6 are \SI{13.2}{\milli\electronvolt} and \SI{15.7}{\milli\electronvolt}, which match the values in \cite{graf2004}.  Experimental results show a peak responsivity of $R_\mathrm{p}= \SI{8.6}{\milli\ampere\per\watt}$ at a wavelength of $\lambda = \SI{84}{\micro\meter}$ corresponding to the first peak in Fig.~\ref{fig:wavefunctions}, which agrees well with the simulated value of $R_\mathrm{p} = \SI{9.4}{\milli\ampere\per\watt}$. Furthermore, a relatively small red shift of the simulation results in comparison to the experimental data can be seen. This might be explained by the highly sensitive processing technology and can give small variations in layer thicknesses and thus variations in the position and energy of the quantized states of the heterostructure. The determined extraction efficiency $p_\mathrm{e} = 0.52$ is in good agreement with the estimated escape probability of $p_\mathrm{e}\approx 0.5$ \cite{graf2004}.
\label{chp:simulation}
\section{Conclusion}
In summary, a sensitive and precise simulation tool for THz QCD performance characterization and optimization is introduced. We applied our modeling approach to an existing THz design and obtained good agreement with the published experimental results. Further extensions of our model towards the characterization of noise dependent quantities such as the detectivity  giving an measure for the signal-to-noise ratio are in progress.
\section{Acknowledgments}

The authors acknowledge financial support by the European
Union's Horizon 2020 research and innovation
programme under grant agreement No 820419 -- Qombs
Project ''Quantum simulation and entanglement engineering
in quantum cascade laser frequency combs'' (FET
Flagship on Quantum Technologies), and by the German
Research Foundation (DFG) within the Heisenberg program
(JI 115/4-2).

\bibliographystyle{IEEEtran}
\bibliography{literature}

\end{document}